\documentstyle[11pt]{article}

\def\be{\begin{equation}}
\def\en{\end{equation}}

\begin{document}
\begin{titlepage}
\baselineskip = 25pt
\begin{center}
{\Large\bf GALAXY CLUSTERS
AND MICROWAVE BACKGROUND ANISOTROPY}

\vspace{.5 cm}
{\bf Vicent Quilis, Jos\'e M$^{\underline{\mbox{a}}}$. Ib\'a\~nez
and Diego S\'aez}\\
\small
Departamento de Astronom\'{\i}a y
Astrof\'{\i}sica. Universidad de Valencia.\\
46100 Burjassot (Valencia), Spain.\\
\footnotesize
e-mail: (Decnet) 16444::saez, (Internet) diego.saez@uv.es\\
\end{center}

\vspace {2. cm}
\normalsize
\begin{abstract}

Previous estimates of the microwave background
anisotropies produced
by freely falling spherical clusters
are discussed. These estimates are
based on the Swiss-Cheese and
Tolman-Bondi models. It is proved that these models give only
upper limits to the anisotropies
produced by the observed galaxy clusters.
By using spherically symmetric codes including pressureless matter and
a hot baryonic gas, new upper limits are obtained.
The contributions of the
hot gas and the pressureless component to the total
anisotropy are compared. The
effects produced by the pressure are proved to be negligible;
hence, estimations of the cluster anisotropies based
on N-body simulations are hereafter justified.
After the phenomenon of
violent relaxation,
any realistic rich cluster can only produce small
anisotropies with amplitudes
of order $10^{-7}$. During the rapid process of
violent relaxation,
the anisotropies produced by nonlinear clusters
are expected to range in the interval $(10^{-6},10^{-5})$.
The angular scales of these anisotropies are discussed.

\end{abstract}


\end{titlepage}
\section{Introduction}

Cosmological overdensities on supercluster scales are pancake-like;
nevertheless, galaxy clusters cannot be considered
as planar structures; clusters are similar to ellipsoids, which
become quasispherical in some cases (Coma cluster).
This shape suggests the use of spherical symmetry as an
approximating condition.
This symmetry strongly reduces the computational cost
with respect to the general tridimensional (3D) case, but
the resulting evolution is too fast.
This problem with the evolution of spherical clusters
is a result
of the radial structure of the velocity field.
Even for vanishing initial velocities, gravitational forces
generate a too
rapid infalling radial motion.
As it is
proved in Section 4, the fast evolution of the
spherical model leads to an overestimate of the
nonlinear anisotropies of the Cosmic Microwave Background (CMB)
produced by galaxy clusters.
Even for very accurate spherical models (including hot gas),
the nonlinear anisotropy is always overestimated.

In the last decade, there has been a lot of work on the
CMB anisotropies produced by nonlinear
cosmological structures. Nearby voids, clusters and Great Attractor-like
objects have been considered.
These structures are often modeled by using
spherically symmetric pressureless solutions of the
Einstein equations. Two popular models are based on this kind of
solutions: (i) The Swiss-Cheese (SC) model proposed by Rees \& Sciama (1968)
is based on a matching of three exact solutions. This model involves rather
particular initial
conditions, and (ii) the Tolman-Bondi (TB) model is
based on the solution obtained by Tolman (1934) and Bondi (1947); in
this second case, initial conditions are general.
The initial profiles of the energy density and the velocity
can be arbitrarily chosen (Arnau et al. 1993).
Chodorowski (1991) used a Newtonian version of the
TB model. This version applies when relativistic effects are
negligible and, consequently, it applies in the case of nonlinear
galaxy clusters.
Nottale (1984) used the SC model to predict relative temperature
variations $\Delta T/T \sim 10^{-4}$ for very dense
and fast collapsing objects, but the masses of these objects
are much greater than
that of the richest cluster.
In the case of feasible rich clusters, recent
estimates based on the TB model and its Newtonian version are
presented in Table 1, where D is the distance from the observer
to the cluster and $M$ is the mass inside a sphere of radius
$R_{_{M}}$.
The resulting predictions depend on the
amplitudes (normalization) and the shapes of the initial profiles.
{}From Table 1, it follows that the
predicted $\Delta T/T$ values corresponding
to realistic rich clusters are a few times $10^{-6}$.
These values are below presently
observable levels. The main limitations of the SC and the TB models
are the spherical symmetry condition and
the absence of pressure.

In the case of very large structures, such
as voids and Great Attractor-like objects, the spherical
pressureless model
seems to be acceptable, but in the case of clusters,  the
combined effect of high density contrasts and
radial infalling velocities lead to
an unavoidable collapse
(see below); furthermore, the hot gas content in clusters is around
$10 \%$ and may even reach up to
$30 \%$ of the estimated virial mass (B\"{o}hringer \& Wiedenmann
1991); hence, the hot gas component is not clearly
negligible. The pressure of this component could produce
very large gradients --even shocks-- which could be comparable to the
gradients of the dominant dark component.
Since the nonlinear
anisotropy produced by a cluster essentially depends on
time variations of the spatial gradients of the
total gravitational potential, the contribution of the hot gas
component should not be neglected without any justification.
This component
should be gravitationally coupled to the
pressureless matter in order
to estimate its importance in the calculation of
anisotropies. A spherically symmetric coupling is studied
in this paper. Some conclusions can be extended to the
general nonsymmetric case.

In a previous paper (Quilis, Ib\'a\~{n}ez \& S\'aez 1994),
it was shown that some shocks develop in the
hot gas component of simplified planar 1D structures, the same
could occur in the spherically symmetric case.
The possible formation of shocks strongly motivates
the use of modern {\it high-resolution shock-capturing} techniques;
nevertheless, the use
of these techniques is preferable in
any case, even in the absence of shocks.
Some advantages of
these codes are described by Ryu et al. (1993).

Other estimates of the CMB anisotropy produced by clusters
are based on N-body simulations. Van Kampen \&
Mart\'{\i}nez-Gonz\'alez (1991)
simulated a rich cluster producing
an effect on the order of a few times
$10^{-7}$.
Anninos et al. (1991) considered a distribution of
nonlinear objects but they did not study the effect
produced by a single rich cluster.
The main problems with N-body simulations are: the absence of a
hot gas component and the uncertainties in both the
spectrum and the statistics of the energy density distribution.
A lot of information about clusters is concerned with
the temperature, the density contrast and the
energy radiated by the hot gas component, but these
quantities are not considered in N-body simulations.
This information should be compared with theoretical
predictions given by a suitable model based on the 3D
coupling between the hot gas (described by
modern {\it high-resolution shock-capturing} techniques) and
the pressureless (described by {\it N-body simulations}) component.
This model is complicated. The comparison between its
predictions and the observations (including those related to
the hot gas component) should be very important in order
to simultaneously test the spectrum, the statistics, and all the
usual hypothesis about the composition and properties of
the hot gas component.

In this paper,
the possible production of shocks and the importance of
the hot gas component are studied in the framework of the
spherically symmetric model. Some conclusions can be
extrapolated to the general
3D case.

The anisotropy produced by a distribution
of nonlinear density perturbations is being currently
studied; recently,
Mart\'{\i}nez-Gonz\'alez, Sanz \& Silk (1994) have reported that
these nonlinear anisotropies range from $10^{-6}$ to
$10^{-5}$ on degree angular scales. The main problems
with this kind of calculations arise as a result of
the current uncertainties about the time evolution of both
the spectrum and the statistics
of the nonlinear density distribution.
Given a null geodesic, any nonlinear object located near
this line influences the corresponding microwave photon; hence,
taking into account the existence of nonlinear structures
from the observer position to distances of order
$\sim 10^{3} \ Mpc$ --the beginning of nonlinearity--,
estimates of the total nonlinear anisotropy would
require: (1) N-body simulations in very large boxes including
the observer and all the structures influencing the CMB and (2)
information about
the time evolution of a nonlinear spectrum involving
all the nonlinear scales.
Since such a spectrum and simulations
are not yet available (even if the hot gas component
is neglected), other approaches giving useful {\em indications}
about the anisotropy produced by a realistic distribution
of nonlinear structures are very useful. One of these
approaches is the estimate of the anisotropy produced by
a single nonlinear cluster.

Here, the spherical freely falling
model for cluster evolution is improved by
introducing a hot gas component. The limitations
of this model are pointed out and some applications to the
estimation of CMB anisotropies are
presented and discussed;
furthermore, another spherical model (Section 4.2) is
used in order to find significant upper limits to the anisotropies
produced by nearby virialized clusters.

The main features of our improved spherical model are:
(i) The clusters are assumed to be nonlinear
structures formed by a
pressureless component --cold dark matter plus point-like galaxies--
and a hot rarified gas emitting in the X--band,
(ii) both components evolve
in the total
gravitational field created by themselves, and (iii)
modern {\it high-resolution shock-capturing} techniques are
used in order to solve
the partial differential equations governing the
evolution of the system.

Hereafter, $t$ stands for the cosmological time, $t_0$ is
the age of the Universe, $a(t)$ is the scale factor.
$\dot{X}$ stands for the derivative of the function X
with respect to the cosmological time. Function $\dot{a}/a$ is denoted by H.
Hubble constant is the present value of H; its value in units of
$100 \ Km \ s^{-1} \ Mpc^{-1}$ is $h$.
The background is flat.
Velocities are given in units
of the speed of light.

The plan of this paper is as follows: In Section 2, our numerical
code is described.
In Section 3, it is proved that any spherically symmetric
model produces a too fast cluster evolution.
Relevant upper limits to the nonlinear
anisotropies produced by galaxy clusters are
shown in Section 4. A discussion about the evolution of
the hot gas component is presented in Section 5,
and the main conclusions are summarized and
discussed in Section 6.

\section{Basic equations and numerical code}

Our model consists of a hot baryonic fluid and a pressureless one,
both fluids are gravitationally coupled.
The hot component is described as a fluid with pressure,
while the dark matter and the point-like galaxies are
two components of an unique pressureless fluid. The
evolution of each of these fluids is described by
their corresponding system of hydrodynamical equations.
Hereafter,
$\rho_{_{T}}$ is the total energy density.
$\rho_{b}$, $\rho_{_{PL}}$ and
$\rho_{_{B}}$ stand for the
energy densities of the hot gas, the pressureless matter
and the
background (critical density), respectively; the same subscripts are used
for the velocities, the density contrasts and any other
quantity defined for each fluid.
The subscript $i$ stands for initial conditions.
All the density contrasts are
defined with
respect to the background energy density, for
example, the baryonic density contrast of the hot gas is
$\delta_{b}=(\rho_{b}-\rho_{_{B}})/\rho_{_{B}}$.
The pressure is denoted $p$, $\epsilon$ is
the specific internal energy,
$T_{7}$ is the temperature in units of $10^7$ K,
and $L_{1}$ is the luminosity produced by an sphere
of radius $h^{-1} \ Mpc$ centred at the point
where the cluster is.

\subsection{Basic equations}

For small enough spatial scales (see below), each of
the gravitationally coupled fluids obeys
the following Newtonian equations:\\
\begin{equation}
\frac{\partial \delta}{\partial t} + \frac{1}{a}\nabla\cdot (1+ \delta)
{\vec {v}} = 0
\end{equation}
\begin{equation}
\frac{\partial {\vec v}}{\partial t} + \frac{1}{a}
({\vec {v}}\,\cdot\, \nabla){\vec {v}}
+ H{\vec {v}} = - \frac{1}{\rho a}\nabla p - \frac{1}{a} \nabla \phi
\end{equation}
\be
\frac{\partial E}{\partial t} + {1 \over a}\nabla \, \cdot \,[(E+p){\vec v}]
=-3H(E+p) -H\rho{\vec v}^2 -\frac{\rho{\vec v}}{a}\nabla{\phi} - \Lambda
\en
${\vec v}=a(t)\frac{d{\vec r}}{dt}$ ,
${\vec r}$, $\Lambda(\rho_{b},T)$, and
$\delta$ being the peculiar velocity, the Eulerian dimensionless
coordinates,
the cooling rate,
and the density contrast of the fluid, respectively.
$E=\rho \epsilon +
\frac {1}{2} \rho {\bf v}^{2}$ is the addition of the
internal and the kinetical energy densities.
Quantities without subscripts correspond to a generic fluid.
The equations governing each particular fluid have the
form (1)-(3), but
the subscripts mentioned in Section 2
must be included in each case.
The total energy density contrast
$\delta_{_{T}}$ is the source of the peculiar gravitational
potential $\phi(t,{\bf x})$, which satisfies the following
equation
\begin{equation}
\nabla^2\phi = \frac{3}{2} H^2 a^2 \delta_{_{T}} \ ,
\end{equation}
this potential is involved in the
differential equations of each fluid (gravitational coupling).
Only pressure gradients and gravitational
forces act on the system.

The Newtonian description given by Eqs. (1)-(4)
applies if the following
conditions are satisfied (Peebles 1980):
a) The inhomogeneity size is much smaller than
the causal horizon size; thus, background curvature is negligible,
and velocities are much smaller than $c$, and
b) no strong local gravitational fields are present. These
conditions make unnecessary a relativistic approach. In all our applications
we have verified that the above conditions are satisfied.

In the spherically symmetric case,
Eqs. (1)-(3) can be easily written as follows:
\begin{equation}
\frac{\partial {\vec u}}{\partial t} + \frac{\partial {\vec f}({\vec u})}
{\partial r}={\vec s}({\vec u})
\end{equation}
\noindent
the vector of unknowns ${\vec u}$ being
\be
{\vec u} = [ \delta , m, E]
\en
\noindent
where $m=(\delta +1)v$.
The vector-valued function ${\vec f(\vec u)}$ (the fluxes) is
\be
\vec f(\vec u) = \left [ \frac{m}{a} ,
 \frac{m^2}{(\delta + 1)a} + \frac{p}{a\rho_{_{B}}}, \frac{(E+p)m}{a
(\delta + 1)} \right ]
\en
\noindent
and the sources ${\vec s(\vec u)}$ are
\begin{eqnarray}
\vec s(\vec u) = \left[ -\frac{2m}{ar}\right.
,& - &\frac{(\delta +1 )}{a}
\frac{\partial}{\partial x}\phi
 - Hm - \frac{2m^2}{ar(\delta+1)},  \nonumber\\
& & \left. \, \,  - 3H(E+p)
 - \frac{\rho_{_{B}}H m^2}{(\delta + 1)} - \frac{m\rho_{_{B}}}{a}
\frac{\partial\phi}{\partial r}-\frac{2(E+p)m}{ar(\delta+1)}
-\Lambda\right]
\end{eqnarray}

Written in this way, we have displayed the conservative character
of the system, in the sense of Lax (1973). Its hyperbolic property
was already pointed out in Quilis, Ib\'a\~nez \&
S\'aez (1993, 1994). Thus, the above system is
a one-dimensional (1D)
{\it hyperbolic system of conservation laws} with
sources. In order to solve numerically these kind of systems,
powerful tools have been developed, the so-called
modern
{\it high-resolution shock-capturing} techniques (see below).

The hot gas component located inside the clusters is a fluid obeying
Eqs. (5)-(8). An equation of state of the form
$p_{b}=(\gamma -1)\rho_{b}\epsilon$ is assumed.
The pressureless component is another fluid obeying the same equations.
The equation of state is $p_{_{PL}}=0$. The peculiar
gravitational potential involved
in the equations governing the evolution of both fluids is the
same; it is the total peculiar gravitational field produced by the
total density contrast $\delta_{_{T}}= (\rho_{b}+ \rho_{_{PL}}
- \rho_{_{B}})/ \rho_{_{B}}$
and, consequently, this potential satisfies the following
1D spherical version of Poisson's equation:
\begin{equation}
\frac{1}{r^2}\frac{\partial}{\partial r} \left[ r^2 \frac{\partial \phi}
{\partial r}\right]
= \frac{3}{2} H^2 a^2 \delta_{_{T}} \ \ .
\end{equation}
At each instant,
the term $\nabla\phi$ is computed as $\nabla\phi=G\frac{M}{R^2}$; where
$R=ar$ and
$M=\int 4\pi \delta_{_{T}}\rho_{_{B}} R^2 dR$ is the total peculiar mass
located inside a sphere of radius $R$ at time $t$.

\subsection{Some details about our code}

We have built up a hydro-code based on a
modern {\it high-resolution shock-capturing}
method.
Our code involves the "minmod" cell reconstruction
(it is a version of the
MUSCL algorithm derived by van Leer, 1979),
Roe's prescription
for evaluating the numerical fluxes (Roe 1981),
and a second order Runge-Kutta algorithm
for advancing in time. It can be proved that these elements
set out a global second order accurate
algorithm.
Reader interested in
details can be addressed to Quilis, Iba\~{n}ez \& S\'aez (1994).

Some comments
about the time steps and the spatial grids used in
this paper are now presented.
The time step used in our numerical integrations
is chosen to be
the smallest of the following times:

1. The Courant time $\Delta t_{c}$,  the
minimum of the steps $\Delta t_{cj}$ defined by
\be
\Delta t_{cj} = CFL_1 \frac{\Delta r}{|\lambda_1(r_{j+{1\over 2}})
- \lambda_3(r_{j-{1\over 2}})|}
\en
where $j$ labels the cell and $j+{1\over 2}$ is the interface between
the cells $j$ and $j+1$. $\lambda_{1}$ and $\lambda_{3}$ are
the minimum and maximum characteristic speeds (see
Quilis, Iba\~{n}ez \& S\'aez 1994).
$CFL_{1}$ is a correction factor  whose
value is experimentally fixed in order to obtain the best results.
Correction factors range
in the interval ($0,1)$.
In our computation, the factor $CFL_{1}$ has varied from 0.6 to 0.9.

2. The dynamical time
\be
\Delta t_{d}= CFL_2 \sqrt{\frac {3 \pi^{2}}{4 \rho_{_{T}}}}
\en
where $\rho_{_{T}}$ is the maximum total energy
density appearing in the
previous time iteration. Typical values of the factor
$CFL_2$ are of the order $10^{-4}$.

The  first epoch of the evolution is governed by
the time step $\Delta t_{c}$. During the ulterior very
nonlinear epoch, that is, when
$\delta_{_{PL}}$ reaches large values, the most restrictive time
step is $\Delta t_{d}$.

In all the calculations displayed in this paper, a geometric spatial
grid with 400 cells is used.
It has been verified that a grid with 800 cells
does not lead to physically significant
differences with respect to the 400 cells grid.
Hence, this number of cells warranties
the convergence to the solution.

Our hydro-code
passed successfully the
standard shock tube tests, both in the Newtonian and relativistic 1D cases
(Mart\'{\i}, Ib\'a\~nez \& Miralles 1990, 1991;
Marquina et al. 1993). Pure
cosmological tests were presented in Quilis, Ib\'a\~nez \&
S\'aez (1993, 1994).

\section{Cluster model}

In this section, the initial conditions for cluster evolution
are fixed in such a way that the resulting
cluster appears to be similar to the observed ones; however,
the evolution of these structures is not realistic.
A very rich Abell cluster is simulated from a particular choice
of the initial conditions.

Radio observations at $21 \ cm$ show that the level
of neutral hydrogen
located inside clusters is very low (Gunn-Peterson test).
In the framework of a cold dark matter model (Cen \& Ostriker
1992),
the high degree of ionization of the hot baryonic gas cannot be produced
by shocks, bremsstrahlung or free-bound radiation;
hence, this ionization must be assumed to be an initial
condition.
The high amount of iron in the hot
gas requires that a large fraction
of this comes from galaxies in bursts and winds (see B\"{o}hringer
\& Wiedenmann 1991 and references cited therein). On account of
these considerations,
the hot gas is assumed to be mainly formed by
high energetic particles produced by stellar evolution --winds,
supernovae-- and by processes in active galactic nuclei.
This part of the gas and the residual primordial gas
(which is not confined inside galaxies during galaxy formation)
are strongly ionized by ultraviolet radiation emitted by
stars,
the resulting gas remains ionized until present time.
It is also assumed that the formation of the ionized gas
and its localization in the intergalactic space occurs
during a cosmologically short period; thus, this gas is
assumed to be instantaneously formed at a
given initial redshift $z_{i}$ (the feasibility of this
assumption is discussed in Section. 5).

It is assumed that galaxies trace mass; thus, the initial
density of galaxies is proportional to that of
dark matter. Since galaxies were formed in the primordial
baryonic gas and they afterwards produced the main part
of the ionized component observed in clusters, the
initial density of
this component is assumed to be proportional to that of
galaxies; hence, all the initial densities are assumed to be
proportional to the total energy density. Density
contrasts are accordingly related among them.

Initial velocities are assumed to be identical for
the three components. The dark matter and the point-like
galaxies are pressureless components obeying the
same equations; hence, their velocities are
identical at any time. As a result of the absence of pressure,
these velocities can be obtained by using
the spherical freely falling solution of the Newtonian
hydrodynamics equations (Peebles 1980).
This solution admits an arbitrary density contrast.
The chosen velocities correspond to vanishing
decaying modes. Finally, since the hot gas
is mainly formed
inside galaxies (and galaxies are formed in the initial
primordial gas), all the hot gas component participates
--at the beginning-- of the same
motions as the point-like galaxies; during evolution, the
hot gas and the pressureless component evolve in a different way and,
consequently,
their velocity fields become different.

The initial profile of the total
density contrast is chosen to be:
\be
\delta_{_{T}}(R,t_i)= \frac{\delta_{_{Ti}}}
{1+\left(\frac{R}{R_V}\right)^{1.8}} \ \ ,
\en
where $\delta_{_{Ti}}$ is the amplitude of
$\delta_{_{T}}(R,t_i)$, and
$R_V$ is the radial distance at which the initial density contrast
reduces to one-half of its amplitude $\delta_{_{Ti}}$.
This profile gives suitable present profiles for galaxies.

There is a set of free parameters in the initial conditions.
The values of these parameters must be chosen in such a way that
the simulated and the observed clusters be comparable.

The free parameters are: the initial redshift $z_i$,
$\delta_{_{Ti}}$, $R_V$,
the initial specific internal energy
$\epsilon_{i}$, and
the ratio between the amplitudes of the hot gas and
the total energy densities.

As the observations
in the X-band and the observations of the Sunyaev -Zel'dovich (1980)
effect suggest, the hot gas component is very rarified.
This gas is mainly formed by protons, electrons, and
helium nuclei. It is highly ionized.
The abundance of protons and He nuclei are assumed to be
of 9 to 1 in number. The resulting gas is treated as a monoatomic gas
(adiabatic coefficient $\gamma=5/3$)
with an averaged atomic weight calculated from the above
standard abundances (Sherman 1982).

The temperature evolution of the rarified hot gas is only
sensitive to the Compton cooling and the
thermal Bremsstrahlung. As a result of the small
density of neutral elements, other interactions like
recombinations and ionizations are not important; hence,
the term $\Lambda$ --involved in some equations of Section
2.2-- can be written as follows: $\Lambda = \Lambda^C + \Lambda^{Br}$,
where $\Lambda^C$ and $\Lambda^{Br}$ are the contributions of
the Compton cooling and the thermal Bremsstrahlung, respectively.
According to Umemura \& Ikeuchi (1984) $\Lambda^C$ and $\Lambda^{Br}$ are:
\be
\Lambda^C=5.4\times10^{-36}(1+z)^4
n_e T  \, \, \, \, \, (erg\, cm^{-3}\,  s^{-1})
\en
and
\be
\Lambda^{Br}=1.8\times10^{-27}n_eT^{1\over 2}(n_{_{HII}} + 4n_{_{HeIII}})
\, \, \, (erg\,  cm^{-3}\, s^{-1})
\en
where $n_e$, $n_{_{HII}}$ and $n_{_{HeIII}}$ are the number density of
electrons, protons, and helium nuclei, respectively.

The main observational features of galaxy clusters are
the following:  The density of electrons --in units of $cm^{-3}$--
ranges in
the interval ($10^{-3}, 10^{-2}$)
(Sunyaev \& Zel'dovich 1980) and the total mass within
a radius of $1.5h^{-1} \ Mpc$ is $\sim 10^{15} \ M_{\odot}$.
These data are only compatible with
$\delta_{b}$ values ranging in the interval ($10^{2},10^{3}$).
The radius of the core is $R_{c} \sim
0.2 h^{-1}  \, Mpc$,
$R_{c}$ being the distance at which the
present energy density reduces to one-half of its maximum value
(Peebles 1994). The temperature of the
hot gas component, $T_{7}$, and the luminosity, $L_{1}$, in units of
$erg/s$
range in the intervals
($2,10$) and ($10^{43},3\times 10^{45}$), respectively
(B\"{o}hringer 1991).
The present density contrast of the galaxy distribution has
the form (12).

For the following initial conditions:
$z_{i}=7$, $\epsilon_{i} =5 \times 10^{-6}$,
$\delta_{_{Ti}}=0.26$, $R_{_{V}}=0.6h^{-1} \ Mpc$, and $\rho_{bi}=
0.2\rho_{_{Ti}}$,
Table 2 shows the features of the resulting structure at the
times $0.94t_{0}$, $0.96t_{0}$, and $0.98t_{0}$.

At time $0.96t_{0}$, the resulting structure looks like a
very rich Abell cluster (perhaps a too rich cluster, but see
below for a justification of this choice); hence, our spherically symmetric
model can reproduce the main features of realistic clusters;
nevertheless, the evolution of the chosen cluster
is not admissible. At time $0.98t_{0}$,
the densities, the luminosity, and the temperature are too great,
while the size is too small; hence, an unavoidable collapse
is developing. At times smaller than $0.94t_{0}$, the densities and the
temperature become too small and the size becomes too large.
The existence of rather stable clusters located between
$z \sim 1$ to $z \sim 0$ is not
compatible with the spherical symmetry condition.
This occurs because the pressureless component has a radial
infalling motion accelerated by gravity
and, consequently, this component fast collapses, forcing
the collapse of the subdominant hot gas component, and
making the clusters
very unstable structures.
As it is proved in the next section, any quickly
evolving  spherical
model (including the model of this paper
plus the SC and TB models)
leads to overestimates of the anisotropies produced by
clusters.
In spite of these facts, the spherical model can be appropriately
used in order to get significant upper limits for the
nonlinear anisotropy.

Figure 1 shows the profiles of
the hot gas and the pressureless density contrasts
at time $0.96t_{0}$. No shocks have appeared in the hot gas
component. Although this component is gravitationally
dragged by the rapid freely falling pressureless component,
pressure has not produced spatial gradients
larger than those of the pressureless fluid;
hence, the presence of pressure
does not seem to be
important in order to compute cluster anisotropies
(see Section 4.1 for a quantitative verification of this
statement). As
discussed in the introduction, the
study of the hot gas component should be important in
order to study the evolution and normalization
of 3D realistic clusters,
but it does not seem to be directly relevant in the calculation
of nonlinear anisotropies. Since the largest gradients and the
shocks are expected to be favored by the
rapid infalling induced by the spherical symmetry condition,
the above conclusion about the
importance of the hot gas component in the
calculation of nonlinear anisotropies can be extended to
the case of realistic clusters. This important fact
justifies the computations of nonlinear anisotropies based on
N-body simulations (excluding the hot baryonic gas).
The evolved profile of the pressureless density contrast
has the form (12), hence,
the predicted profile of point-like galaxies is
compatible with observations.

Since shock formation is expected to be more feasible in the case
of the richest clusters, we have preferred the study of a very
rich one (perhaps too rich); thus the absence of shocks in
the chosen case ensures the absence of these phenomena
in the case of any realistic rich cluster.

Objects similar to Abell clusters can be simulated at any time
by using our freely falling model. Admissible times would
range from
$z \sim 1$ to $z \sim 0$. The evolution appears to be
wrong in any case.

\section{CMB anisotropies}

Any overdensity located between the observer and his last
scattering surface produces: (1) a Sachs-Wolfe effect, (2)
A Doppler effect, which appears as a result of the peculiar
velocity produced
by the overdensity on the last scattering, (3) a second
Doppler effect due to the peculiar velocity induced on
the observer, (4) a Sunyaev-Zel'dovich effect produced
by the interaction between the CMB photons and the free
electrons of the overdensity, and (5) a gravitational
nonlinear effect.

In the case of a nonlinear cluster, the effects (1) and (2) are
negligible because these structures are very far from the
last scattering surface. The effect (3) is only relevant when
the cluster is located very near to the observer (Virgo cluster),
it has an exactly dipolar form. The effect (4) is the dominant
one in the case of a single object, its estimate is
easy and its detection has been claimed by
Uson \& Wilkinson (1988) and Birkinshaw (1990).
The total Sunyaev-Zel'dovich effect produced by a realistic
distribution of clusters is a matter of current study.
The study of the effect (5) is the main subject of this section.

When the nonlinear effect (5) dominates,
the anisotropy is (Mart\'{\i}nez-Gonz\'alez, Sanz \& Silk
1990):

\be
\frac{\Delta T}{T} \sim
-2\int_{e}^{o} {\vec {\nabla} \phi . ({\vec x},t)} {d{\vec x}}
\sim 2\int_{e}^{o} \frac {\partial{\phi}(\vec x,t)}{\partial{t}} dt  \ ,
\en
\noindent
where $\phi$ is the potential involved in the line element

\be
ds^2=-(1+2\phi) dt^2 + (1-2\phi)a^2 \delta_{ij}dx^idx^j  \ ;
 \en
\noindent
if this line element and the energy-momentum tensor of a perfect fluid
are introduced in the Einstein equations and the powers and
products of the potential and its gradients are neglected, the
Newtonian equations (1)-(4) are obtained; hence, the function $\phi$
involved in Eqs. (15) and (16)
is the Newtonian gravitational potential created by
the cluster.

The integral involving $\frac {\partial{\phi}(x,t)}{\partial{t}}$
points out the importance of the time evolution of the
gravitational potential. Any overestimate of
$\frac {\partial{\phi}(x,t)}{\partial{t}}$ leads to an
overestimate of $\Delta T/T$.

The integral involving the
spatial gradients of the potential $\phi$
is the most appropriated in order
to be evaluated by using our numerical methods.
These gradients are directly involved in the
differential equations describing the cluster evolution;
hence, when these equations are numerically solved,
the spatial gradients are directly calculated
at the nodes of the spatial grid and at each time step; afterwards,
the gradients can be calculated at arbitrary positions and
times by using suitable interpolations.

The integrals involved in Eq. (15) must be carried out along
each null geodesic from the emitter (e) to the observer (o). The
emitter is located on the last scattering surface.
The equations of these geodesics can be derived
in the background (at zero order).

In the spherically symmetric case, a direction of observation
(a null geodesic) is defined
by the angle $\psi$ formed by the line of sight and the line
pointing towards de cluster centre.
The origin of coordinates is located in the
cluster centre and the position of the observer is
fully determined by its radial distance to the
origin.

\subsection{Previous upper limits}

If the spherical cluster simulated
in Section 3 is located at $70h^{-1} \ Mpc$ from the observer,
the resulting amplitude of the nonlinear anisotropy is
$\mid \Delta T/T (\psi=0) \mid = 1.2 \times 10^{-5}$ (see Fig. 2).
The CMB
photons passed near this cluster at time $0.96t_{0}$,
when  the structure had the features of a very rich Abell cluster
(see Table 2). The masses inside three spheres
of radius $1.5h^{-1} \ Mpc$,
$2h^{-1} \ Mpc$ and
$4h^{-1} \ Mpc$ are $0.95 \times 10^{15}h^{-1} \ M_{\odot}$,
$2.0 \times 10^{15}h^{-1} \ M_{\odot}$ and
$6.2 \times 10^{15}h^{-1} \ M_{\odot}$, respectively.
These data facilitate comparisons with the predictions
of Table 1.
Let us discuss in detail some of these comparisons. We begin with
Panek's predictions (1992). In the
model described in the second row of Table 1, there is a total mass
$M=5.7 \times 10^{15} h^{-1} \ M_{\odot}$ inside a
sphere of radius $4 h^{-1} \ Mpc$, this mass is very similar to that
of the cluster considered in this section; however, the amplitude
of the density contrast corresponding to our cluster (see
Table 2) is much greater
than that of Panek's model ($\sim 674$); as a result of this
discrepancy,
the amplitude of the anisotropy obtained by Panek (1992)
is one-half of our amplitude.
Let now consider Chodorowski's estimates (1991).
In the cases presented in the rows 3 and 4 of
Table 1, the masses located
inside spheres of radius $\sim 2h^{-1} \ Mpc$
are smaller than
the mass corresponding to our model
($2.0 \times 10^{15}h^{-1} \ M_{\odot}$)
and,
consequently, our results cannot be directly compared
with those of Table 1; nevertheless, an indirect comparison can
be established taking into account that
the amplitude of the anisotropy computed
by Chodorowski (1991) scales with the mass like $\sim M^{3/2}$
(see Nottale 1984), this fact can be
easily verified from the data exhibited in
the rows 3 and 4 of Table 1. The mentioned scaling leads to
the conclusion that, in Chodorowski's models
(1991), the amplitude of the anisotropy
coresponding to $M=2.0 \times 10^{15}h^{-1} \ M_{\odot}$
is $\sim 1.14 \times 10^{-5}$; this value is
in very good agreement with our estimate. After these comparisons,
we can conclude that our
prediction seems to be compatible with those of
Chodorowski (1991) and Panek (1992). The differences between
our amplitude and those of the above authors appear as
a result of the differences in the
normalization and slope of the density and velocity profiles.
Our normalization takes
into account the features of the hot gas component (see
Section 3).
Previous amplitudes and the amplitude computed in this section
have been overestimated (as the $\phi$ evolution).
They are upper limits.

The spatial gradients of the peculiar gravitational potentials produced
by the hot gas and the dark matter components have been
separately calculated. These gradients have been used in order
to compute the anisotropy produced by each of these components
according to Eq (15).
In Fig. 2,
the total temperature contrast $\Delta T/T$ and the part
of this contrast produced by the pressureless matter
are shown. The difference ($\sim 18.5 \%$ of the total effect)
is produced by the hot gas component.
Since the $20 \%$ of the total mass is contained in the
hot gas  and this gas
produces the $\sim 18.5 \%$ of the total anisotropy,
the contribution of the hot baryonic component
to $\Delta T/T$ is very
similar to the expected contribution ($\sim 20 \%$) corresponding to
the same proportion of pressureless matter; hence,
the effect of the pressure is not important. This conclusion
is in good agreement with
the absence of shocks and large gradients in the density
profile corresponding to the hot gas (see Fig. 1)

\subsection{New upper limits}

In order to obtain more stringent upper limits to the
nonlinear anisotropy produced by clusters, it must be taken
into account that: (a)
Any overestimate
of the time evolution of the gravitational potential
$\phi$ leads to an upper limit
for $\Delta T/T$, and (b) the smaller the
differences between the true evolution and the
overestimated one, the smaller and more stringent
the resulting upper limit. After virialization, the
upper limits of Section 4.1 are not stringent enough
because
they are based on too great overestimates
of $\frac {\partial \phi} {\partial t}$; stringent limits corresponding
to this period are obtained in this section.

In realistic clusters, the particles (point-like galaxies and
small gravitationally bounded amounts of cold dark matter)
do not move radially; after the process
of violent relaxation (Lynden-Bell 1967), these particles reach
quasistable orbits in the total gravitational field of the structure
and the inertial forces almost compensate the gravitational
ones; thus the fall
towards the centre becomes very slow and the clusters
appear to be highly stable structures
undergoing modest changes.
The shape of the system can be quasispherical.

After violent relaxation, the total density contrast, $\delta_{_{T}}$,
of the
resulting cluster has been estimated to be $\sim 2 \times 10^{2}$
(B\"{o}hringer and Wiedenmann 1991); hence, according to
our definition of the contrasts (see Section 2), in a model
containing $20 \%$ of baryonic hot gas, the corresponding
$\delta_{b}$ value is $\delta_{b}=  (\rho_{b}/ \rho_{_{B}})-1
= (0.2 \rho_{_{T}} / \rho_{_{B}})-1 = 0.2 \delta_{_{T}}-0.8
  \sim 40$. These
estimates are based on a very simple spherical homogeneous
model. In more realistic models based on nonuniform
density profiles (see Fig. 1 and Eq. 12),
the distribution of mass is not homogeneous and the
expected amplitudes  of $\delta_{_{T}}$ and $\delta_{b}$
should reach values greater than
those predicted by the homogeneous model. On account of
these facts, it is hereafter assumed
that, after virialization,
the amplitude of $\delta_{b}$ is $\sim 100$. This conservative
assumption is also compatible with the limits on
$\delta_{b}$ discussed in Section 3, which are
based on observations.

As discussed in B\"{o}hringer and Wiedenmann (1991),
all the clusters do not virialize at the same redshift;
according to these authors, the virialization finishes
at a certain time depending
on the initial conditions defining the protocluster in the
linear regime. A conservative hypothesis compatible with
observations and theoretical considerations is that
any cluster undergoes virialization at a redshift $z_{vir} \leq 1$.

During the violent relaxation,
the gravitational field undergoes rapid variations
and, consequently, the upper limits based on the
spherically symmetric model of Section 3
are much better than
in the subsequent slowly evolving period. After violent
relaxation, the situation is very different; the
evolution is not known, but two
possibilities can be imagined: (a) the cluster
tends to a stationary state and, (b) the
cluster tends to collapse but at a rate which is much slower than
during the relaxation. In case (a), the evolution of
virialized clusters at $z \ll z_{vir}$
would be very slow. In
case (b), these clusters would evolve faster than in case (a)
as a result of the progressive instability of the system.
In order to find upper limits for the anisotropy produced by
virialized clusters at $z \ll z_{vir}$, a certain evolution of kind (b)
is appropriately simulated.

According to the observations,
all the galaxy clusters --from $z \sim 1$ to $z \sim 0$--
have a hot baryonic component with density contrast $\delta_{b}$
ranging from $10^{2}$  to $10^{3}$; hence,
the fastest admissible evolution of kind (b) would produce a present
cluster with $\delta_{b} \sim 10^{3}$ starting from the
value $\delta_{b} \sim 10^{2}$ corresponding to $z_{vir}$
(see above).
Any quantity being bounded by the observations --as
the total mass inside a sphere of $1.5h^{-1} \ Mpc$-- can be
used to define overestimated evolutions in the same way
as $\delta_{b}$.
Any overestimated evolution is fictitious. It does not describe the
true evolution of the clusters. It is only used with the
essential aim of finding upper limits to the nonlinear
anisotropy. This kind of evolution can be produced by
fictitious suitable forces. The use of fictitious forces is
not a newness in Cosmology, they are introduced in the
{\it Adhesion model} (Gurvatov, Saichev \& Shandarin
1989) with the essential aim of simulating
structure formation; nevertheless, in our case, the
role of the fictitious forces is more modest than in the
adhesion case.

An evolution of kind (b) can be simulated by means of
a fictitious force producing a partial cancelation of the
gravitational one. A perfect cancelation leads to stationarity.
The strongest the cancelation, the slowest the evolution.
If the fictitious force $\vec {F}_{f}$ is assumed to be proportional to
the gravitational one ($\vec {F}_{f} = (1- \zeta) \vec {\nabla} \phi$, $
0< \zeta <1$),
the inner regions becomes more stable than the outer ones, where
gravitational forces are greater; thus, some kind of accretion
on the core is simulated. This core is not fully stable
because the
gravitational force increases as a result of
the accretion, but the instability
does not produce a too rapid collapse as in the absence of
any compensating action.
The residual force leads to a certain degree of
evolution.
The value of the parameter $\zeta$
can be numerically determined --in each case-- by assuming a final
$\delta_{b}$ value of order $10^{3}$; thus a too fast
evolution is simulated (at least for $z \ll z_{vir}$).
$\zeta=0$ corresponds to stationarity.

In practice, the computations are carried out as follows:
(1) the model described in Section 3 is used in order to
obtain suitable density contrasts ($\delta_{b} \sim 10^{2}$)
at $z = z_{vir}$.
These contrasts plus vanishing velocities are the
initial conditions for the subsequent evolution,
in which, the pressure is neglected
and a fictitious antigravity force is introduced.
These conditions are easily included in the codes
described in Sections 2 and 3.
The parameter $\zeta$ is fitted in order to get
suitable overestimates of the evolution; namely,
suitable final values $\delta_{b} \sim 10^{3}$.

The case described in Fig. 3 corresponds to a
cluster with contrasts $\delta_{b} \sim 100$ and
$\delta_{b} \sim 1200$ at redshifts $z=z_{vir}=1$ and $z=0$,
respectively. At the small redshifts corresponding to the assumed
locations of the cluster ($z < 0.1$),
$\delta_{b}$ is greater than $10^{3}$.
A parameter $\zeta \sim 5.5 \times 10^{-3}$ has
been necessary in order to simulate this evolution. This quantity
gives a qualitative idea about the high degree of
stability being necessary in order to explain the
observations about clusters. Since the gravitational forces are
very strong, the main part of them
must be cancelled in order to keep acceptable
levels of accretion (variations of $\phi$).
The simulated object has a total mass $\sim 0.6 \times 10^{15}
h^{-1} \ M_{\odot}$
inside
a radius of $1.5h^{-1} \ Mpc$. It is a rich cluster.
Fig. 3 shows the nonlinear anisotropy
for three positions of the system. In these positions,
the distances from
the cluster centre to the observer are
$25h^{-1} \ Mpc$,
$100h^{-1} \ Mpc$, and
$300h^{-1} \ Mpc$.
The amplitude of the
nonlinear anisotropy is $\sim 4 \times 10^{-7}$ in all the
cases, but the angular scales consistently change.

The results displayed in Fig. 3
have been obtained in the case $z_{vir} = 1$. This high value of
$z_{vir}$ has been arbitrarily chosen. Values of
$z_{vir}$ smaller than 1
have been also considered, but the results corresponding to
these values are not
displayed in Figures by the sake of briefness. These results lead to
the following conclusion: If
a rich cluster is normalized
in the same way as in the case $z_{vir}=1$
and located at $z \ll z_{vir} < 1$,
the resulting anisotropy
is $\sim 4 \times 10^{-7}$ for any admissible choice of $z_{vir}<1$.
This anisotropy is very similar to that
predicted by van Kampen and Mart\'{\i}nez-Gonz\'{a}lez by
using N-body simulations.
Since $\delta_{b}$ increases from $10^{2}$ to $\sim 10^{3}$
in all the cases, the fictitious evolution corresponding
to $z_{vir}=1$ is slower than that of the cases
$z_{vir}<1$ and, consequently, the resulting anisotropy
should be an increasing function of $z_{vir}$; since this
dependence on $z_{vir}$ is not found,
the fictitious evolution --although faster than the true
evolution-- should be negligible in all
the cases. In order to verify this possibility,
the gravitational
force has been completely canceled
($\zeta=0$) and the anisotropy has been calculated again;
as expected,
the results are indistinguishable from those presented in
Fig. 3. Two main conclusion can be obtained:
(1) At $z \ll z_{vir}$, any admissible evolution
is too slow to produce a significant contribution to $\Delta T/T$
and (2) the resulting $\Delta T/T$ values obtained in our
computations
are produced by a non-evolving spherical cluster located at
various positions.  The time delay of
the CMB photons crossing a nonlinear stationary structure
produces the estimated anisotropy. This effect was first described
by Rees an Sciama (1968).

{}From the above results, the following conservative conclusion
can be obtained: Any rich cluster
produces an effect of a few times $10^{-7}$ at $z \ll z_{vir}$.
The angular scale of this anisotropy
depends on the position of the cluster (see Fig. 3 for
the angular scales corresponding to three locations).
Any realistic distribution of virialized clusters cannot produce
relevant effects ranging in the interval
$(10^{-6}, 10^{-5})$.

Nearby
clusters located at distances smaller than $300h^{-1} \ Mpc$
($z < 0.1$) can only produce relevant anisotropies if
they are undergoing violent relaxation.
Taking into account that the distances between clusters are
of various tens of Megaparsecs, this kind of nearby objects
can only appear in some isolated directions.
The effect of these clusters could be separately considered
after detection in observational surveys; hence, we are
hereafter concerned with non virialized clusters
located at $z>0.1$, which would produce
anisotropies on scales smaller than $\sim 1^{\rm o}$. Are these
anisotropies cosmologically relevant?. Some comments about
this question are worthwhile.

During the epoch of violent relaxation, cluster
evolution is faster than in the subsequent epoch; hence, the
most important anisotropies must be produced during
relaxation. In this epoch,
the upper limits given by the freely infalling
model of Section 3 are much more stringent than at
$z<z_{vir}$.
These limits can be obtained in the same way as
in Section 4.1, but the epoch of infalling and the location
of the cluster must be appropriately fixed near $z_{vir}$.
The resulting upper limits are of the same order as in
Section 4.1; therefore,
the anisotropies arising during relaxation
could reach values near
$10^{-5}$ for the richest clusters.
If a cluster is located at $z \sim 1$,
the angular scale of the corresponding anisotropy is about
$10'$, while the angular scale corresponding to
$z=0.1$ is $\sim 1^{\rm o}$; hence, the total anisotropy produced
by clusters
undergoing violent relaxation at different redshifts
$0.1<z_{vir}<1$ is a superposition of angular scales ranging
from $10'$ to $1^{\rm o}$.
According to this discussion,
we claim that the main part of the nonlinear effect (5)
should be produced
during a brief period ($0.1<z<1$) corresponding to
violent relaxation of clusters; in other words,
only the clusters
located in a narrow interval of redshifts would
produce significant anisotropies and each of these cluster would
only produce anisotropy during the short period of violent
relaxation.
These facts could guide
and simplify
future estimates based on N-body simulations.

The anisotropy produced by a distribution of clusters
cannot be calculated from that of a single cluster,
but amplitudes at the
level of
$10^{-5}$ or $10^{-6}$
(see Mart\'{\i}nez-Gonz\'alez, Sanz \& Silk 1994),
on angular scales between $10'$ and $1^{\rm o}$
are suggested by the above arguments. More work
is necessary
in order to see if the currently observable level
($\sim 10^{-5}$) is reached.

\section{Some comments on the hot gas component}

If the pressure of the hot gas component is not
neglected from $z_{vir}$ to $z=0$ --in order to obtain
temperatures and luminosities--, and
the compensating antigravity force is introduced --in
order to stabilize the system--, the above
conclusions about anisotropies do not change; however,
it has been verified that the Compton cooling and the thermal
Bremsstrahlung strongly cool the system. Starting from
initial admissible values of the
temperatures and luminosities, the evolved
values become too low.
The clusters are radiating energy during a long period
--from $z_{vir}$ to the redshift defining the cluster
location-- and, consequently, a too strong
cooling is produced. This effect decreases as $z_{vir}$
decreases.
The same effect is expected in
realistic nonsymmetric cases (although the stabilization
is not produced by fictitious forces).
This problem is a result of the hypothesis that all
the hot gas component was generated and distributed
at a certain initial
redshift (see Section 3). The hot gas component must
be continuously ejected from galaxies and a substantial
part of this component must arrive to
the intracluster medium
after $z_{vir}$. This is a necessary ingredient
of future nonsymmetric simulations.

\section{Conclusions}

Cluster evolution
has been proved to be slower than the evolution
predicted by spherical models; on account of this fact,
these models can only be used in order to
obtain upper limits to the anisotropy produced by
galaxy clusters.

The CS and
TB models lead to very fast
evolutions, which only give stringent enough upper limits
to $\Delta T/T$
during the brief period of violent relaxation.
Nottale (1984), Chodorowski (1991), Panek (1992) and
S\'aez, Arnau \& Fullana (1993) presented some estimates based on
these models (see Table 1).

Appropriate spherical models evolving faster than the
observed clusters are used in order to obtain
upper limits to the microwave background anisotropies
produced by some clusters.
In the case of a virialized nearby clusters ($0 \leq z \leq 0.1$),
the angular scale of the anisotropy is larger than $1^{\rm o}$
and its amplitude is a few times $10^{-7}$.
The anisotropy produced by a realistic distribution of
these structures is too small to be detected.
Nonlinear clusters undergoing virialization in the
period $0 \leq z \leq 0.1$ cannot be very abundant, these
objects can be only observed in isolated directions (see
Section 4.2), they lead to small local spots with amplitudes of a
few times $\sim 10^{-6}$ (perhaps near $10^{-5}$ for very
rich clusters).
Only clusters undergoing violent relaxation at
redshifts $0.1<z<1$ can produce
significant anisotropies on large regions. The total effect of all
these clusters is a superposition of anisotropies on angular scales
ranging from $\sim 1^{\rm o}$ to $\sim 10'$.
Upper limits based on the spherical
freely infalling model prove that
this anisotropy is smaller than $10^{-5}$ (likely
a few times $10^{-6}$).
Realistic cluster models are necessary in order
to improve these conclusions.

In order to simulate realistic clusters, it would be
desirable a suitable
coupling between a 3D code based on
modern {\it high-resolution shock-capturing} techniques
and a N-body code. The first code would describe
the evolution of the hot gas component and the
second one would describe the pressureless matter.
Both components must evolve in their total gravitational
field.
Here, it has been verified that the hot gas component
must be gradually generated.
In our opinion, the most interesting
feature of these realistic models is the study of
the temperature and the luminosity of the hot baryonic
component, whose evolution
--for given cosmological spectra and (or)
statistics--
can be compatible (or not) with observational data.
Furthermore, these future models could be basic in order
to estimate the Sunyaev-Zel'dovich effect produced
by the hot gas located inside clusters. Fortunately,
according to our estimates, neglecting
the hot gas component (N-body simulations) must lead to an
accurate enough computation of the nonlinear gravitational anisotropy
produced by clusters; of course, a model
including this component would give more
accurate estimates of these anisotropies.

\vspace{1 cm}

\noindent
{\it Acknowledgements}. This work has been
supported by the  Generalitat Valenciana (grant GV-2207/94)
and Spanish DGICYT (grant
PB91-0648). Anonymous referee is acknowledged for
useful suggestions.
Calculations were carried out in a VAX 6000/410 at the Instituto de
F\'{\i}sica Corpuscular, in a IBM 30-9021 VF at the Centre de
Inform\`atica de la Universitat de Val\`encia, and in a DECstation 2100
at the Centro T\'ecnico
de Inform\'atica del Consejo Superior de Investigaciones
Cient\'{\i}ficas. V. Quilis thanks to the Conselleria d'Educaci\'o i
Ci\`encia de la Generalitat Valenciana for a fellowship.

\vspace{1 cm}

\noindent
{\large{\bf References}}\\
\\
Anninos P., Matzner R.A., Tuluie R. \&
Centrella J., 1991, ApJ, 382, 71\\
Arnau J.V., Fullana M.J., Monreal L. \&
S\'{a}ez D., 1993, ApJ, 402, 359\\
Birkinshaw M., 1990, in {\em The cosmic microwave
background: 25 years later}, eds. N. Mandolesi \& N. Vittorio,
77.Dordrecht:Kluwer\\
B\"{o}hringer H., 1991,
in {\em Lectures Notes in Physics}, New insights
into the Universe, eds. Mart\'{\i}nez V.J., Portilla M. \& S\'{a}ez D.,
Springer-Verlag\\
B\"{o}hringer H. \& Wiedenmann G., 1991,
in {\em Lectures Notes in Physics}, New insights
into the Universe, eds. Mart\'{\i}nez V.J., Portilla M. \& S\'{a}ez D.,
Springer-Verlag\\
Bondi H., 1947, MNRAS, 107, 410\\
Cen R. \& Ostriker J, 1992, ApJ, 393, 22\\
Chodorowski M., 1991, MNRAS, 251, 248\\
Gurvatov S., Saichev A. \& Shandarin S.F., 1989, MNRAS,
236, 385\\
Lax P., 1973,
in {\em "Regional Conference Series Lectures in Applied
Math."}. SIAM, Philadelphia\\
Lynden-Bell D., 1967, MNRAS, 136, 101\\
Marquina A., Mart\'{\i} J.M$^{\underline{\mbox{a}}}$.,
Ib\'a\~{n}ez J.M$^{\underline{\mbox{a}}}$.
\& Miralles J.A. \& Donat R., 1992, A\&A. 258, 566\\
Mart\'{\i} J.M$^{\underline{\mbox{a}}}$.,
Ib\'a\~{n}ez J.M$^{\underline{\mbox{a}}}$.
\& Miralles J.A., 1990, A\&A. 235, 535\\
Mart\'{\i} J.M$^{\underline{\mbox{a}}}$.,
Ib\'a\~{n}ez J.M$^{\underline{\mbox{a}}}$.
\& Miralles J.A., 1991, Phys. Rev. D43, 3794\\
Mart\'{\i}nez-Gonz\'alez E., Sanz J.L. \&
Silk J., 1990, ApJ, 355, L5\\
Mart\'{\i}nez-Gonz\'alez E., Sanz J.L. \&
Silk J., 1994, In press\\
Nottale L., 1984, MNRAS, 206, 713\\
Panek M., 1992, ApJ, 388, 225\\
Peebles P.J.E., 1980,
{\em The Large Scale Structure of the Universe}.
Princenton University Press\\
Peebles P.J.E., 1994,
{\em Principles of physical Cosmology}.
Princenton University Press\\
Quilis V., Ib\'a\~{n}ez J.M$^{\underline{\mbox{a}}}$.
\& S\'aez D., 1993,
Rev. Mex. Astron. Astrof\'{\i}s. 25, 117\\
Quilis V., Ib\'a\~{n}ez J.M$^{\underline{\mbox{a}}}$.
\& S\'aez D., 1994, A\&A. 286, 1\\
Rees M.J. \& Sciama D.W., 1968, Nat, 217, 511\\
Roe P.L., 1981, J. Comp. Phys. 43, 357\\
Ryu D., Ostriker J.P., Kang H. \& Cen R.,
1993, ApJ, 414, 1\\
S\'aez D., Arnau J.V. \& Fullana M.J., 1993, MNRAS,
263, 681\\
Sherman R.D., 1982, ApJ, 256, 370\\
Sunyaev R.A. \& Zel'dovich Ya. B., 1980, A\&A. 18, 537\\
Tolman R.C., 1934, Proc. Natl Acad. Sci., 20, 169\\
Umemura M. \& Ikeuchi S., 1984, Prog. Theor. Phys. 72, 47\\
Uson J.M. \& Wilkinson D.T., 1988, in {\em Galactic
and extragalactic Astronomy}, eds. G.L. Verschuur \& K.I.
Kellermann, Springer-Verlag\\
van Kampen E. \& Mart\'{\i}nez-Gonz\'alez E., 1991,
in Proceedings of Second Rencontres de Blois "Physical
Cosmology", ed. A. Blanchard, L. Celnikier, M. Lachieze-Rey
\& J. Tran Thanh Van (Editions Frontieres), p 582\\
van Leer B., 1979, J. Comp. Phys. 32, 101\\

\newpage
\noindent
{\Large{\bf Figure captions}}\\
\\
{\bf Figure 1.} Density contrasts --at time $0.96t_{0}$-- of
the hot baryonic gas (continuous line) and the pressureless
component (dashed
line) as functions of the
radial distance R in units of $h^{-1} \ Mpc$.\\
\\
{\bf Figure 2.} Dotted line is a
plot of $-(\Delta T/T) \times 10^{5}$ as a function
of the angle $\psi$ (in degrees) for the rich cluster
described in Table 2
located at $70h^{-1} \ Mpc$ from the observer.
Dashed line displays the contribution of the pressureless
component to the
total effect (dotted line).\\
\\
{\bf Figure 3.} Plot of $-(\Delta T/T) \times 10^{7}$ as a function
of the angle $\psi$ (in degrees) for the quasistable
rich cluster of Section 4.2 located at the distances from
the observer displayed
inside the panel.

\newpage

\begin{table}
\begin{center}
{\bf Table 1.} Previous estimates of $\Delta T/T$.\\

 \begin{tabular}{ccccc}
    Authors &  $D$ & $R_{_{M}}$ & $M$ & $\Delta T/T$\\
        &($h^{-1} \ Mpc)$ &$(h^{-1} \ Mpc)$ &
        $(\times 10^{15} \ h^{-1} \ M_{\odot})$ & $(\times 10^6)$\\
   Panek & $100$  &  $4$ & $1.8$ & $-1.0$\\
   Panek & $100$   &     $4$ & $5.7$ & $-6.0$\\
   Chodorowski & $\sim 70$  &     $2$ & $0.77$ & $-2.6$\\
   Chodorowski & $\sim 70$  &     $2$ & $1.54$ & $-7.7$\\
\multicolumn{5}{c}{}\\
\end{tabular}
\end{center}
\end{table}

\vspace{3 cm}

\begin{table}
\begin{center}
{\bf Table 2.} The spherically symmetric model.\\

 \begin{tabular}{cccccc}
       time  &  $\delta_{b}$ & $\delta_{_{PL}}$ &
$L_{1}$
& $T_{7}$ & $R_{c}$ \\
	& & & $(\times 10^{44} \ erg/s)$ & & $(\times h^{-1} \ Mpc)$\\
    $0.94t_{0}$  &  $110$ & $510$ & $0.37$ & $1.2$ & $0.75$\\
    $0.96t_{0}$   &     $520$ & $3220$ & $4.6$ & $3.3$ & $0.23$\\
    $0.98t_{0}$  &     $4000$ & $1.9 \times 10^{5}$
    & $42.0$ & $120.0$ & $0.02$\\
\multicolumn{6}{c}{}\\
\end{tabular}
\end{center}
\end{table}

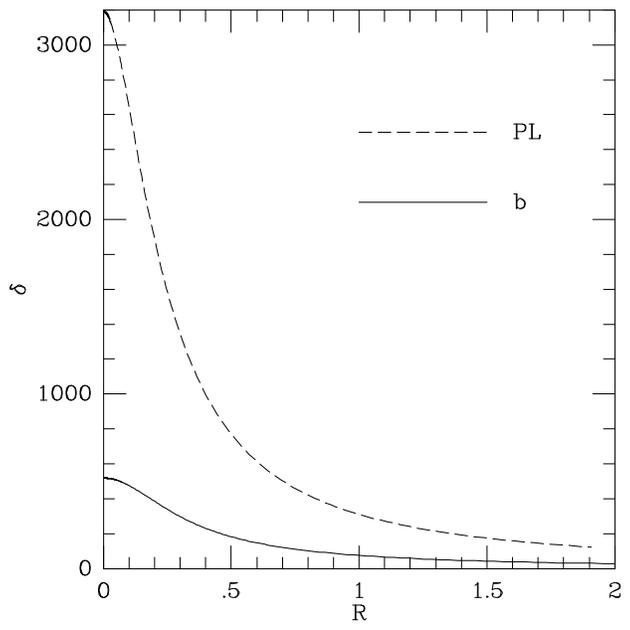
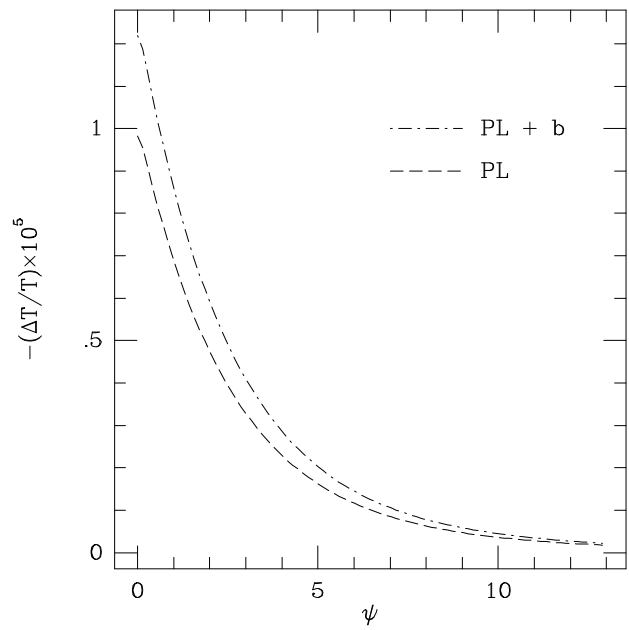
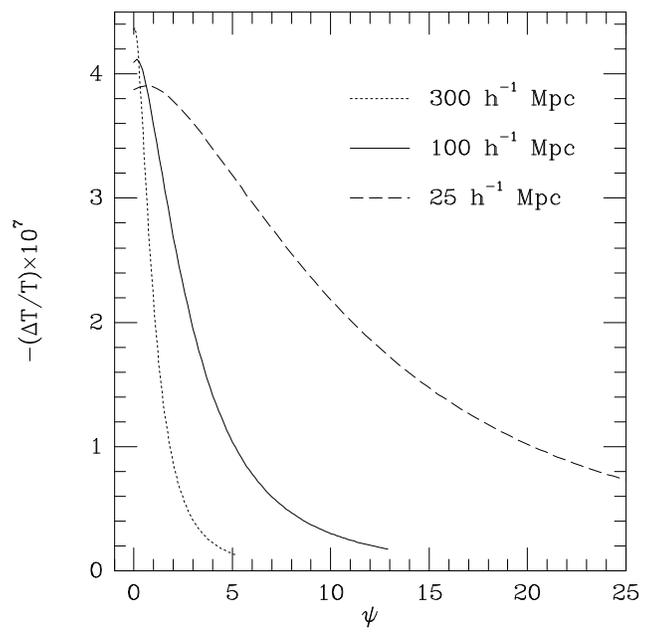
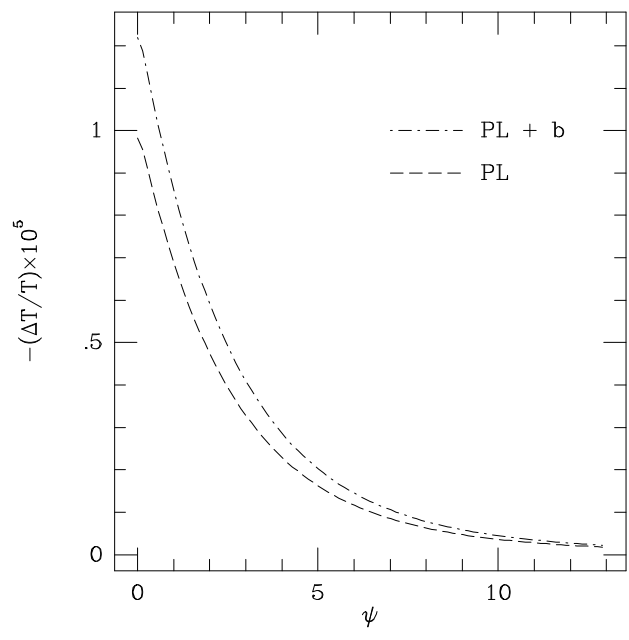
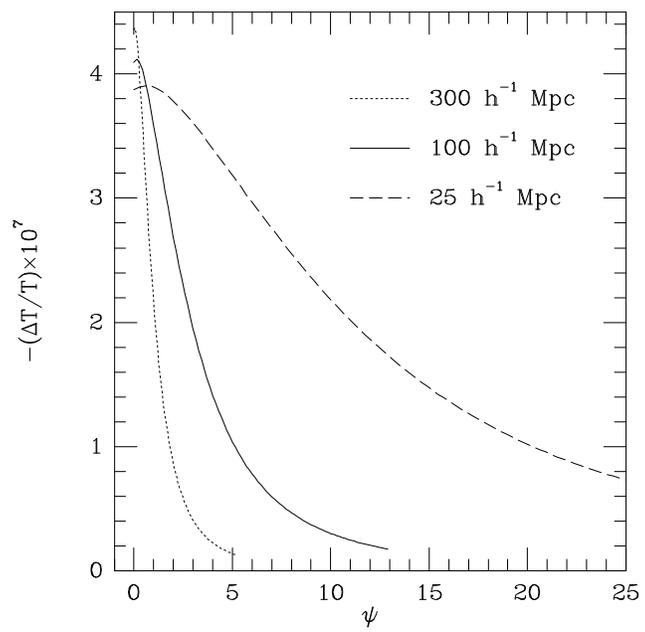
\end{document}